\documentclass[12pt, letterpaper]{article}

\usepackage{amsmath,amssymb}
\usepackage{cite}
\usepackage{fancyhdr}
\usepackage[top=1in, bottom=1in, left=1in, right=1in]{geometry}
\usepackage{graphicx}
\usepackage{hyperref}

\numberwithin{equation}{section}
\date{}

\begin{document}
\title{{\rm\footnotesize \qquad \qquad \qquad \qquad \qquad \ \qquad \qquad \qquad \ \ \ \ \ \                      RUNHETC-2024-13,    UT-WI-04-2024
}\vskip.5in    Holographic Inflation, Primordial Black Holes and Early Structure Formation\\  Submitted to 2024 Gravitation Research Foundation Essay Contest May 13, 2024}
\author{Tom Banks\\
NHETC and Department of Physics \\
Rutgers University, Piscataway, NJ 08854-8019\\
E-mail: \href{mailto:tibanks@ucsc.edu}{tibanks@ucsc.edu}
\\
\\
Willy Fischler (corresponding author)\\
Department of Physics and Texas Cosmology Center\\
Weinberg Institute, Center for Theory\\
University of Texas, Austin, TX 78712\\
E-mail: \href{mailto:fischler@physics.utexas.edu}{fischler@physics.utexas.edu}
}

\maketitle
\thispagestyle{fancy} 

\begin{abstract}  Evidence has accumulated that there are supermassive black holes (SMBHs) in the centers of most galaxies, and that these were formed in the very early universe by some as yet unknown process.  In particular, there is evidence\cite{silketal} that at least some galaxies formed as early as $10^8 - 10^9$ years after the Big Bang host SMBHs.  We suggest that the holographic model of inflation, whose dark matter candidates are primordial black holes carrying a discrete gauge charge, which originated as a small subset of the inflationary horizon volumes in the very early universe, can provide the seeds for this early structure formation.  Aspects of the model pointed out long ago suggested an early era of structure formation, with structures dominated by dark matter.  The additional assumption that the dark matter consists of discretely charged black holes implies black hole dominance of early structures, which seems to be implied by JWST data.

\normalsize \noindent  \end{abstract}


\vspace{1cm}

\vfill\eject
\section{A Speculation About the Seeds For Early Galaxy Formation}

The holographic model of inflationary cosmology\cite{holocosmrevised} began as an attempt to construct a finite quantum model of the earliest moments of the universe, matching on the flat FRW universe with equation of state $p = \rho$ in the hydrodynamic limit.  Later, we realized that it could incorporate a smooth termination in a model of de Sitter (dS) space with a finite dimensional Hilbert space.  The additional insight that localized objects in dS space are constrained subspaces of the dS Hilbert space, with the property that subsystems in those subspaces remain independent for long periods of time\footnote{This may seem somewhat obscure when stated abstractly, but the explicit matrix models of\cite{bfm} show exactly how it is possible.}, leads to a formulation of inflationary cosmology that resolves the Trans-Planckian problem of conventional inflation.

We can explain this, without too much detail, by pointing out that our model\cite{mannelli} of the flat FRW universe with 
\begin{equation} a(t) = \sinh^{1/3}(3t/R_I) , \end{equation} ended up with a system whose density matrix was
\begin{equation} \rho = \frac{e^{-L_0}}{{\rm Tr}\ e^{- L_0} } , \end{equation} where $L_0$ is the Virasoro generator of a (cut-off) $1 + 1$ dimensional CFT, with central charge proportional to the area of the inflationary horizon.  This is exactly the same as the Carlip\cite{carlip}-Solodukhin\cite{solo} ansatz for the density matrix of a black hole of the same area. In the cosmological context this is the density matrix of a causal diamond along any geodesic in the FRW universe, for times $> R_I$.  We remind the reader that
\begin{itemize}
\item In Holographic Space Time Models, the background space-time is viewed, following Jacobson\cite{ted95} as a hydrodynamic approximation to the underlying quantum theory. Homogeneity and isotropy are built into the model by choosing the same time evolution operator $U(t)$ along each geodesic in the background, and by having the variables transform under the rotation group that preserves a geodesic, and the time dependent Hamiltonian be invariant.
\item Flatness follows from the dynamics of this particular model at all times, as a consequence of insisting that the covariant entropy bound is saturated.
\end{itemize}

In order to build a realistic model of cosmology, containing localized objects, we allow the horizon to expand after some number of inflationary e-folds, but constrain the initial conditions such that individual horizon volumes do not interact for a long time.  The mechanism for this involves two separate features.  The first is that the interaction between localized subsystems is mediated by exciting degrees of freedom that are set to zero by the initial conditions.  A model for this is a matrix model with single trace interactions in which localized degrees of freedom are blocks in a matrix that has been set to be block diagonal by the initial conditions.   The second feature is the Milne redshift.  The interactions which turn the frozen degrees of freedom on and off are all located on the stretched apparent horizon and have a natural time scale which is the horizon radius.   If the horizon expands more rapidly than the time it takes to equilibrate the frozen variables, then the localized objects will survive.  Since their internal dynamics is identical to that of black holes, we argued that the slow roll era ends with a dilute gas of black holes.  In the explicit quantum model the Milne redshift is implemented by writing the time dependent Hamiltonian at any time as a sum of terms corresponding to contributions from different time-like trajectories in the conformally rescaled Casini-Huerta-Myers\cite{CHM} coordinates defined below, with red shift factors read from that metric.  Excitations closer to the geodesic are defined as those that evolve more rapidly in the quantum Hamiltonian.  

Given a flat FRW metric, Einstein's equations define a pressure and energy density $p,\rho$, and if $p + \rho \geq 0$ we can invent an "inflaton" field such that the metric is sourced by the inflaton.  This field has no relation to any field in low energy effective field theory around a Minkowski vacuum of string/M-theory, so none of the constraints of\cite{vafa} apply to it.  The only constraint on the slow roll metric is that the expansion rate during slow roll is faster than the scrambling rate which would equilibrate the localized black holes with the rest of the degrees of freedom on the apparent horizon.  This is
\begin{equation} \epsilon \equiv \frac{\dot{H}}{H^2} > C {\rm ln} [(M_P/H)]^{-1} . \end{equation}
$C$ is a constant characteristic of a particular fast scrambling system and is not easy to determine from the formula for the Hamiltonian.  
Combined with the observational constraint from the CMB that
\begin{equation} \epsilon^{-1} (H/M_P) \sim 10^{-5} , \end{equation}
this leads to the estimate 
\begin{equation}\epsilon \sim 0.1\ if\ C\ is\ o(1).   \end{equation}   As we'll see, this is consistent with data, because our prediction for the tensor to scalar ratio differs from that in conventional inflation models.  

There is no semi-classical picture of the slow roll era in terms of particles or a quantum field theory.  It is defined by a quantum mechanical model in which causality is imposed by making the Hamiltonian time dependent so that degrees of freedom close to a geodesic only couple to those far from it after a sufficient proper time has elapsed. Geometrically, one is working in coordinates\cite{CHM} where, at a coarse grained level,
\begin{equation} ds^2 = a^2 (t) \frac{T}{2}(\cosh t + \cosh x)^{-2} (- dt^2 + dx^2 + \sinh^2 x d\Omega^2) . \end{equation} Here $T$ is the proper time since the beginning of the universe and $a(t)$ the scale factor.  We call this coarse grained because after the horizon begins to expand, and we have distinct non-interacting quantum systems, each of which behaves like an inflationary horizon volume, we cannot place these systems on geodesics that are a Planck length apart for all time, as we did during the earliest moments of the universe.   When the slow roll era ends we have, on FRW time slices, a collection of black holes that must be separated by more than their Schwarzschild radii, in order to be consistent with treating them as approximately non-interacting.  So "homogeneity" exists only on this larger length scale, and even here it is violated by the statistical fluctuations in the black hole entropy/mass.   We can calculate the size of those fluctuations from the Carlip-Solodukhin ansatz, and the tensor to scalar ratio $r$ is determined\cite{tbas} by the entropy formula for Kerr black holes.  One finds that $r$ scales like $\epsilon^2$ rather than $\epsilon$.   The reason for this is that, according to the C-S ansatz  $\frac{\delta H}{H} $ is independent of $\epsilon$, whereas in field theoretic inflation it scales like $\sqrt{\epsilon}$.   The fluctuation calculation is done by assuming that the classical Einstein equations, linearized around some slow roll metric, give a coarse grained hydrodynamic description of our quantum model, with stochastic initial conditions set by our assumption that we have a dilute gas of black holes, with relation between the fluctuations in $\frac{\delta M}{M}$ and $\delta{L}{L}$ given by the Kerr formula.  The mass distribution of the black hole fluctuation is smoothed to a horizon sized Gaussian and the spacing between black holes is left as an input parameter in the fit, as is, of course, the slow roll metric.  

At a given proper time $T$ since the beginning of the universe, the variables of our quantum model are two dimensional quantum fermion fields (with a UV cutoff), labelled by $ T \times T + 1$ matrix indices.  The initial conditions are such that the square $ T \times T$ matrices bilinear in these are block diagonal, with a number $K$ of $R_I \times R_I$ blocks and one block of size $ T - K R_I $.  The Hamiltonian is a single trace in matrix bilinears.  According to\cite{hilbertbundles} the bilinear term is just a sum of free Dirac Hamiltonians, while the quartic term is a certain abelian Thirring interaction.  One term in the Hamiltonian involves the full $T \times T$ matrix and evolves the system with a characteristic time scale $T$.   In addition to this there are terms of the same form involving only a single $R_I \times R_I$ block matrix.  Each of those is associated with a particular $x$ value in the CHM coordinates of the diamond at time $T$.  The one associated with the geodesic at $x = 0$ evolves in Planck units, while each other value of $x$ is assigned the appropriate red shift factor.   We will not go into details of the quantum model of the slow roll era here.  It is unlikely that they will lead to testable predictions, and we mention the model only to demonstrate that an explicit system incorporating the principles we've outlined, exists. 

The slow roll era is followed by an era dominated by a marginal dilute gas of black holes.  We use the word marginal because any state with localized excitations is constrained, and is {\it a priori} a less probable initial condition than one which leads to the maximal entropy universe with $a(t) = \sinh^{1/3} (3t/R_{final}) , $ for some choice of the final dS radius.  Thus, the most probable state with localized excitations is one where the post inflationary gas of black holes is just dilute enough to escape this fate.   This means that the black hole number density on a surface of fixed FRW time is 
\begin{equation} n_0 = \frac{C_0}{R_{IBH}^3} , \end{equation} where $C_0$ is something like $10^{-2} - 10^{-3}$ and $R_{IBH}$ is the size of the apparent horizon at the end of inflation.   From this point on, we can follow our cosmology semi-classically.  The dilute black hole gas is a matter dominated universe and density fluctuations grow.  The initial magnitude of the fluctuations is determined\cite{tbas} by our identification of the early inflation phase with a system whose modular Hamiltonian is the cutoff $L_0$ generator of a CFT.  They are of order\footnote{Note the different power of the slow roll parameter, compared to field theoretic inflation models.} $\epsilon^{-1} (R_{IBH}/L_P)^{-1}$ .   Fluctuations go non-linear when
\begin{equation} t \sim (\epsilon R_{IBH} / L_P)^{3/2} L_P , \end{equation}, which is much shorter than the black hole decay time
\begin{equation} t_{decay} \sim g^{-1} (R_{IBH}/L_P)^3 L_P . \end{equation}   $g$ is the number of particle degrees of freedom with mass below the Hawking temperature of the decaying black holes. Primordial structures thus begin to develop in the dilute black hole gas (DBHG) before black hole decay sets off the Hot Big Bang (HBB). Crude estimates\cite{holocosmrevised} put the reheat temperature at $10^{10}$ GeV.

We have also postulated\cite{tbwfremnants} that a fraction $f$ of the inflationary horizon volumes carry a discrete $Z_N$ gauge charge, which implies that their decay will leave over stable remnants of at least Planck mass.   We now note two further complications.  It is likely that each of these remnants is part of a bound cluster, possibly containing other remnants, and certainly containing other decaying black holes.  Furthermore, if there are superpartners in the world, half of the decay products of black holes will actually be massive particles, with masses above $10^3$ GeV.   Many of them will also have gauge interactions that are long range on the scale of the bound clusters.  Thus, the fate of the bound clusters as the black holes decay is a complicated dynamical problem.  Rather than try to solve it we will simply assume that the result is a distribution of primordial black holes, with masses ranging from the Planck mass to black holes of the size of the horizon at the end of the black hole dominated era.   We estimate this to be $\sim 10^{14} - 10^{15} L_P$.   The smaller black holes quickly decay down to Planck scale remnants, leaving no trace but those near the horizon size have Hawking temperatures well below the reheat temperature and are meta-stable for relatively long periods.

 Given this conjecture, the radiation dominated era begins with a sprinkling of near horizon size black holes, along with a collection of stable Planck mass black hole remnants.  Novikov and Zeldovich\cite{novzel} observed long ago that the evolution equation for black holes in an FRW universe with a single component equation of state had a scaling solution in which the black hole size tracks the size of the apparent horizon.  Black holes much smaller than the horizon grow slowly, but in the homogeneous approximation seem to evolve to the scaling solution.  Carr and Hawking\cite{carrhawk} argued that inhomogeneous corrections would prevent the evolution of black holes much smaller than the horizon size to the scaling branch of the equation, and that no black hole solution that is asymptotically FRW can exactly follow the scaling solution.   Of course, once Hawking radiation is taken into account, these holes will also have a probability to decay.  Although there is still an exact scaling solution it is unstable, and generic initial conditions lead to eventual black hole decay,  even in the homogeneous approximation.   However, for our estimate of the horizon size at the beginning of the radiation dominated era, the decay does not begin until the temperature has fallen below at least $10$ GeV.   During this time, lots of interesting things can happen.
 
 The oldest black holes that have been observed\cite{silketal} by the James Webb Space Telescope (JWST) have a mass of about $10^{44} M_P$ and their Schwarzschild radius was the horizon size when the age of the universe was about a second.   So in order to explain black holes of this magnitude we need only hypothesize that some near horizon size black holes, which formed during the black hole dominated era of our cosmology, followed the NZ scaling solution for this fairly short period during the radiation dominated era.  We have no other probes of cosmology at these early times, so this does not seem implausible.  A crucial question, to which we have not yet found an answer, is what determines $10^6$ solar masses as the scale where Carr-Hawking instabilities halt the growth of black holes following the NZ scaling solution.  {\it The determination of the scale at which the NZ scaling solution is halted is the crucial test of our model.}  If a natural explanation of the observed scale of galactic center black holes comes out of this mechanism, we would find it hard to believe that it was incorrect.  A more detailed numerical investigation of the Carr-Hawking equations, within the parameters of our model might lead to an understanding of this length scale.
 
 There is also some evidence that the surprising early galaxies seen by JWST have unusual elongated "banana" shapes\cite{bananas}.  We would like to propose a crude model for such elongated shapes.   Our model of the radiation dominated era consists of a gas of "radiation" with an initial temperature between $ 10^9 - 10^{10}$ GeV.   Half of the radiative species are really massive superpartners of standard model particles with masses in the range $10^3 - 10^5$ GeV.   In addition there is a population of stable Planck mass primordial black holes, and another population of large black holes whose mass grows from somewhere around $10^{14} M_P$ to around $10^{44} M_P$ in the first  second of the radiation dominated era.   Together the two populations of black holes make up the dark matter.  Most of the dark matter consists of stable Planck mass black hole remnants.

 By construction, each of the black holes originated as an inflationary horizon volume moving on a geodesic of the coarse grained FRW model that describes the black hole dominated era of cosmology before fluctuations go non-linear.  It is plausible that after the brief eras of very early structure formation and black hole decay, the resulting swarm of remnant black holes are not moving rapidly in the rest frame of the gas of radiation.   If we now consider a pair of large black holes, the Newtonian escape velocity of the pair is
 \begin{equation} v_{esc} \sim \frac{M_1 M_2 n_L^{1/3}}{M_P^3} , \end{equation} where $n_L$ is the large black hole number density.  Whenever this is larger than the Hubble velocity $H/M_P$, we expect a black hole bound state.   If it has a reasonable angular momentum it will be meta-stable because gravitational radiation is a slow process.  
 
 We want to emphasize that this non-relativistic description of bound state formation is not consistent with causality, but could be a consequence of causal processes.  In HST models\cite{newton} Newton's law arises causally from processes in second order perturbation theory, in which frozen q-bits linking two localized objects in the same causal diamond are excited and de-excited.  This leads to a phase in the transition amplitude for the two objects to enter and exit the diamond at different points on its holographic screen, which can be interpreted semi-classically as arising from a non-relativistic instantaneous potential on FRW time slices. We are using this classical non-relativistic language as shorthand for something that is quite complex to describe in a proper causal and quantum mechanical manner.  
 
 The large black hole bound state is of course immersed in a dense gas consisting of radiation and small black holes, and these will be attracted to the black holes.   But gas on the axis between the two will be pulled in opposite directions, so we should expect a shape with a preferred axis, determined by the initial angular momentum in the large black holes.   The rotation of the black holes will produce a Coriolis force on the matter between them, which will tend to swirl it into a disk shape, but since the motion of the heavy black holes takes place on a much longer time scale, there will be a period where the structures have a large dipole moment.  
 At these early times, it's obvious that the mass of these structures is mostly dominated by the black holes at the ends.   There are indications from the JWST data that very early galaxies have a higher ratio of black hole to normal mass, and this might be the beginning of the explanation for that fact.   Of course, the black holes at the end of the bananas will be absorbing matter and growing all the time, and there are surely frictional processes which could slow the rotation and cause the structure to collapse and the black holes to merge.
 
 It is also probable that many of the large black holes do not pair up into binaries, so these may lead to a qualitatively different type of galaxy.  Perhaps there is some reason that galaxies based on single large black holes take longer to begin star formation, so that they are too dim to be seen in the current JWST observations.  Star formation takes place mostly in regions of high dark matter density.  Near single black holes this will occur mostly near the black hole itself and the dark matter may be absorbed before stars can form.  On the other hand, in the elongated systems, dense clusters of dark matter in the center could light up and will remain stable against the competing pull of opposing black holes.  It's important to estimate the time scale for collapse of elongated systems into more spherical ones in this binary model of early galaxies.   
 
 The observations of very early galaxies go back to a time hundreds of millions of years after the Hot Big Bang, whereas our theoretical considerations are at most valid over the first few seconds.   There is a lot more detailed modeling to be done before one can say that our models have successfully explained the surprising features of the youngest galaxies in the universe.   They do however provide a solid basis for expecting to find a very early universe populated by a collection of large black holes, as well as Planck mass primordial black hole dark matter.  
 
 To summarize:  the HST model of inflationary cosmology leads one to expect an early era of structure formation that ends with a Hot Big Bang set off by the decay of most of the Inflationary Black Holes.  We've also postulated that a certain fraction of the black holes leave over Planck mass remnants, whose stability is guaranteed by a discrete $Z_N$ gauge charge. The temperature of the "radiation" at the beginning of the radiation dominated era is approximately $10^{10}$ GeV.   Our ignorance about particle physics above the scale of $1-2$ TeV leaves open the possibility that much of the "radiation" actually consists of massive particles, and certainly much of it experiences long range forces\footnote{Long range always means of order the apparent horizon.} stronger than gravitation.   So it's entirely plausible that some of the stable black holes have evolved to horizon sized structures at the beginning of the radiation dominated era.
 The horizon size at this time is only about $10^{14} L_P$.  
 
 Such horizon sized black holes will follow the NZ scaling solution until Carr-Hawking inhomogeneities stop their growth.   Since the Schwarzschild radius of even the largest super massive black hole known is miniscule compared to the size of the apparent horizon at any era to which we have direct observational access, we conjecture that the origin of all SMBHs is in fact in this very early era.   We do not as yet understand why the scale of $10^6$ solar masses is the scale around which black hole growth following the NZ solution stops.  
 Bound binaries of such primordial SMBHs will attract the Planck mass black hole dark matter, as well as the baryons and radiation in the universe, and form elongated shapes in the early universe.  It's important to calculate how long these elongated shapes last before collapsing into more spherical profiles.  One would not however expect that most of the early SMBHs formed binaries,  so one would expect another class of early galaxies, centered around single SMBHs.   The present authors do not understand enough about star formation to speculate about why the elongated ones are more easily visible, or whether failure to see a second population of early galaxies rules out a model like this.   One possible explanation relies on the fact that star formation tends to occur in overdensities of dark matter.   For a galaxy formed around a single SMBH overdensities will occur preferably near the central black hole and may be absorbed by it before they can form stars.   For black hole binaries, overdensities can occur on the axis between the two black holes and survive long enough to form stars.  We note that there are alternative models of early prolate galaxies\cite{primack} based on LCDM simulations without SMBHs.  As far as we know there has not yet been any confirmation that the observed elongated early galaxies harbor SMBHs, or that the early galaxies containing SMBHs are elongated.  Thus, our conjecture must survive many observational tests before it can be confirmed.
 
 The holographic model of inflation, combined with the assumption of a discrete gauge symmetry that stabilizes a certain fraction of black hole remnants, is a cosmological model that can account for most of the know facts about the very early universe in terms of a very small number of parameters.   It is based on some general ideas about quantum gravity, is completely finite and does not suffer from "Transplanckian" paradoxes.  It predicts that dark matter takes the form of primordial black hole remnants, which are very hard to detect by anything but their gravitational effects.  In this paper we have shown that it may also provide us with an intriguing explanation of the properties of the very earliest galaxies, which are beginning to be revealed by the JWST.

\vskip.3in
\begin{center}
{\bf Acknowledgments }
\end{center}
 The work of T.B. was supported by the Department of Energy under grant DE-SC0010008. Rutgers Project 833012.  The work of W. F. was supported by the National Science Foundation under grant PHY-2210562.




\end{document}